\documentclass[nologo,11pt,a4paper]{ETHpaper}
\usepackage[square,numbers,sort&compress]{natbib}
\usepackage{caption}
\usepackage{mathtools}

\begin{document}

\newcommand{\mean}[1]{\left\langle #1 \right\rangle} 
\newcommand{\abs}[1]{\left| #1 \right|} 
\newcommand{\ul}[1]{\underline{#1}}
\renewcommand{\epsilon}{\varepsilon} 
\newcommand{\eps}{\varepsilon} 
\renewcommand*{\=}{{\kern0.1em=\kern0.1em}}
\renewcommand*{\-}{{\kern0.1em-\kern0.1em}} 
\newcommand*{\+}{{\kern0.1em+\kern0.1em}}

\newcommand{\RA}{\Rightarrow}
\newcommand{\bbox}[1]{\mbox{\boldmath $#1$}}

\title{How emotions drive opinion polarization: An agent-based model}

\titlealternative{How emotions drive opinion polarization: An agent-based model}

\author{Frank Schweitzer$^{1,2,*}$, Tamas Krivachy$^{1}$, David Garcia$^{2,3}$}

\authoralternative{F. Schweitzer, T. Krivachy, D. Garcia}

\address{$^1$ Chair of Systems Design, ETH Zurich, Weinbergstrasse 58, 8092 Zurich, Switzerland\\
$^2$Complexity Science Hub Vienna, Josefst{\"a}dter Strasse 39, 1080 Vienna, Austria\\
$^3$Section for Science of Complex Systems, CeMSIIS, \\ Medical University of Vienna, Spitalgasse 23, 1090 Vienna, Austria\\
$^*$ Corresponding author: \url{fschweitzer@ethz.ch}
}

\reference{(Submitted for publication)} 
\www{\url{http://www.sg.ethz.ch}}

\makeframing
\maketitle

\begin{abstract}
    We provide an agent-based model to explain the emergence of collective opinions not based on feedback between different opinions, but based on emotional interactions between agents.
    The driving variable is the emotional state of agents, characterized by their valence and their arousal.
    Both determine their emotional expression, from which collective emotional information is generated.
    This information feeds back on the dynamics of emotional states and of individual opinions in a non-linear manner.
    We derive the critical conditions for emotional interactions to obtain either consensus or  polarization of opinions.
    Stochastic agent-based simulations and formal analyses of the model explain our results.
    Possible ways to validate the model are discussed. 
    
  \end{abstract}
\date{\today}

\section{Introduction}
\label{sec:Introduction}

In the past decades, the significance of emotions in opinion formation and decision making has been recognized by the scientific community, and its study is mainly pioneered by the field of behavioral economics and empirical psychology. 
Experimental research on individual behavior show that emotions fuel information sharing \citep{Berger2011}, that emotional arousal drives reactions to become more extreme \citep{Reisenzein1983}, and that emotional states frame the way we process information \citep{Schwarz1991}.
Recent results on observational data analysis suggest that collective opinions and decisions, such as election outcomes or pricing dynamics, can be influenced by  emotional states like collective mood.
A notable example has been the analysis of mood in \texttt{Twitter} and the stock market \citep{Bollen2011,Lachanski2017}, which led to further studies of \texttt{Twitter} emotions in pricing of other assets like Bitcoin  \citep{Garcia2015}.

The modelling of emotions and opinions has mainly focused on their interaction at the individual level, modeling how affective and cognitive mechanisms influence each other \citep{Forgas2008}.
This has left aside the modelling of the interplay between emotions and collective opinions, mainly due to the absence of data to test those models and the technical challenge to formulate and test hypotheses on them.
Our aim in this article is to fill that gap, proposing a computational model of collective opinions and emotions.
Such a model, when designed based on plausible and testable assumptions, can be used as a hypothesis generator to guide future empirical research \citep{Smith2007}.

We apply the principles of a modelling framework of collective emotions in which the collective state arises through interactions via a common information field, and not through one-on-one interactions \citep{Schweitzer2010a}.
The individual dynamics of this model have been calibrated against results from self-reports \citep{Kuppens2010} and empirical data on the dynamics of emotional interaction \citep{Schweitzer2016}.
We model how the opinions of agents are influenced by a common information field accessible to all agents, capturing this way how opinions evolve as a function of shared emotions.
We present analytical and numerical results on this model that constitute stylized facts that can be tested in future empirical studies.

\section{Background}
\label{sec:Background}

Emotions are psychological states of high relevance for the individual that imply cognitive and physiological effects.
They are closely related to our behavior and how we interpret our own actions \citep{Frijda1987}.
Current research in affective science conceives emotions through their measurable components or dimensions, proposing models to quantify and measure them.
The component process model of emotions \citep{Scherer1984} conceptualizes emotions as composed of sequential appraisal processes of the experience of an emotion, such as evaluating if an event is positive or adverse, and parallel processes that include physiological dynamics, action tendencies, and subjective feelings.
The circumplex model of core affect \citep{Russell1980} focuses on emotions on a short timescale in which they do not have a clear target but consume energy in the organism and strongly affect cognition and action tendencies, like verbal expression.
This model quantifies emotions in a dimension of valence (pleasure associated with the emotion) and of arousal (degree of activity induced by the emotion), allowing most emotional experience to be mapped in a circle in this space \citep{Scherer2005}.
While further dimensions are informative of the experience of emotions, like potency and unpredictability \citep{Fontaine2007}, valence and arousal have prevailed as the two major factors to measure short-lived emotional states.
Recent research has modelled emotions as dynamical systems.
For example \citep{Sander2005} modelled emotional fight-or-flight reactions as cusp catastrophes, and DyAffect \citep{Kuppens2010,Lodewyckx2011} calibrated Ornstein--Uhlenbeck dynamics of emotions against empirical data.

The emotions of humans do not exist in isolation and often collective emotional states are triggered or emerge in a crowd.
Collective emotions are defined as emotional states shared by large amounts of people at the same time \citep{vonScheve2013}.
Research on collective emotions, while learning from established works from social psychology and sociology, is still a growing field \citep{vonScheve2014,holyst2016cyberemotions}. 
The hyperlens model of social regulation of emotion is an adaptation of previous models of social factors of emotions to capture collective aspects of emotional life \citep{Kappas2013}.
This model calls research to ``get out of the lab'' and investigate collective emotions in the naturalistic scenarios where they appear.
Studies on shared emotions in collective gatherings have shown the long-term effects of these collective emotions for the feelings of social cohesion and identity of those involved \citep{Paez2015}.
Similarly collective emotions and group-based emotions play key roles in intergroup conflicts \citep{Goldenberg2014}.

The availability of data produced by the digital society motivated the study of collective emotions in online communities and social media.
Collective emotions have been analyzed through sentiment analysis of real-time group chats \citep{Garas2012}, product reviews \citep{Garcia2011}, and of forum discussions \citep{Chmiel2011}.
To understand their emergence and dynamics, the agent-based modelling framework we follow in this article was designed to simulate collective emotions in a variety of online media \citep{Schweitzer2010a}.
This framework was designed to be testable in both controlled experiments and observational analyses of digital traces, following a wider trend of calibrating agent-based models against large-scale datasets \citep{Fortunato2013}.
This modelling framework has been adapted and used to explain \emph{polarization of emotions} in product reviews \citep{Garcia2011}, collective emotions in chatrooms \citep{Garas2012,Tadic2013} and in blogs \citep{Mitrovic2012}, and emotion spreading in the \texttt{MySpace} social network \citep{Tadic2017}.
Further applications of this framework have focused on modelling how bots and dialog systems could drive collective mood in a discussion \citep{Tadic2013b,Rank2013,Skowron2017} and to drive the emotion dynamics of 3D virtual humans in real-time interaction \citep{Ahn2012}.

While such results from data-driven modeling of emotions were quite convincing, recent proposals to formally relate \emph{emotion} dynamics  to \emph{opinion} dynamics have received much less evidence and support.
Opinion dynamics itself is an established field of research and one of the areas, where methods from statistical physics have been successfully applied to model social phenomena \citep{Castellano2009,Schweitzer2018}.
A highly relevant question tackles the emergence of consensus and polarization, motivating  bounded confidence models \citep{Lorenz2007}, information accumulation systems in modular networks \citep{Shin2010}, and studies of the role of biased assimilation and assortativity in the polarization of opinions \citep{Mas2013,Dandekar2013}.

To link emotions and opinion polarization, recent computational models  simply rephrased the dimension of valence as opinion, to then study cusp catastrophes of state changes depending on arousal \citep{Sobkowicz2012,Sobkowicz2013} and on tolerance parameters inherited from bounded confidence models \citep{Sobkowicz2015}.
To date, what is missing is a model that includes both the \emph{fast dynamics} of emotional states and the \emph{slower dynamics} of opinions in an \emph{integrated approach} that can explain the role of emotions in opinion polarization.
This needs to be done based on  principles testable in psychological studies and observational analyses, rather than recasting previous models by simply replacing the terminology of opinions for the one of emotions.
We aim to fill this gap by formulating such model and providing an analysis of its dynamics, opening new research questions for future empirical research.

\section{Agent-based model}
\label{sec:agent-based-model}

\subsection{Modeling emotional dynamics}
\label{sec:quant-emot-dynam}

The main focus of our model is to explain the evolution of opinions based on \emph{emotions}.
Precisely, we do not assume that agents respond to the \emph{opinions} of other agents, directly.
Instead, we assume that the expression of opinions is tightly coupled to the expression of \emph{emotions}, and these emotions in fact influence other agents. 
This allows us to build on our previously developed agent-based framework of emotional influence \citep{Schweitzer2010a}.
It has proven to describe collective emotional states in different social systems \citep{Garas2012,Garcia2011,Tadic2013b,Mitrovic2012}.

\begin{figure}[htbp]
  \centering
  \includegraphics[width=0.45\linewidth]{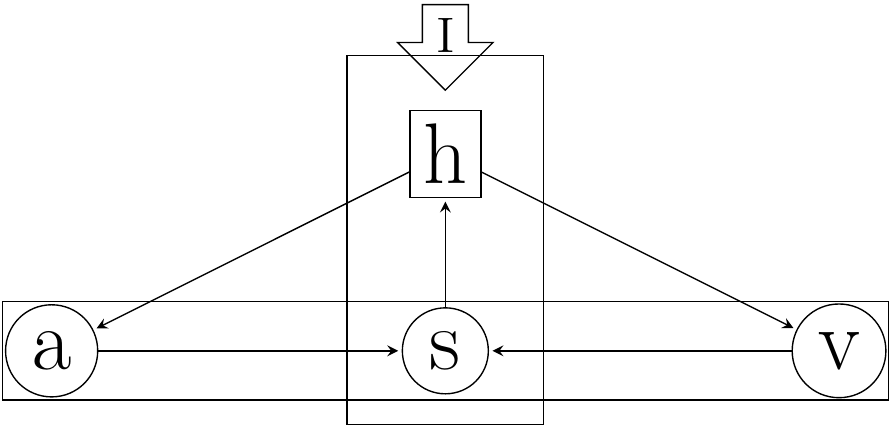}
  \caption{Schematic representation of emotional influence in our agent based model \citep{Schweitzer2010a}. The horizontal layer describes the generation of an emotional expression ($s$) dependent on the valence ($v$) and the arousal ($a$) of the agent. The vertical layer describes the generation an emotional information field ($h$) from such expressions. $I$ denotes possible external events generating emotional information. The emotional field $h$ then influences the valence and arousal of all other agents.
  The formal expressions for these influences are given in Sect. \ref{sec:quant-emot-dynam}.
  }
  \label{fig:schema}
\end{figure}

\paragraph{Emotional information}

Here, we only recap the core dynamics of our agent-based framework of emotional influence, schematically shown in Figure \ref{fig:schema}.
The horizontal layer represents the agent. 
Its emotion is characterized by the two dimensions valence $v$ and arousal $a$. 
Both determine the content of the written expression $s$, which contributes e.g. to an online discussion, as follows:
\begin{equation}
  \label{eq:3}
  s_{i}(t)= s\; \mathrm{sign}\{v_{i}(t)\}\; \Theta\left[a_{i}(t)-\tau_{i}\right]
\end{equation}
Agent $i$ makes an individual contribution $s_{i}$ if its arousal $a_{i}(t)$ exceeds a given individual threshold $\tau_{i}$.
$\Theta[x]$ is the Heavyside function: $\Theta[x]=1$ only if $x\geq 0$.
The threshold $\tau_{i}$ is a random value selected uniformly from an interval $[\tau_{\mathrm{min}},\tau_{\mathrm{max}}]$.
If an agent expresses itself, the emotional content of that contribution results from its valence $v_{i}$ at that time, precisely from the \emph{sign} of its valence, i.e. positive or negative, but the amount of the contribution, $s$, is a parameter equal for all agents. 

The emotional information generated this way by all agents participating is contained in the information field $h(t)$.
It consists of two components, $h_{+}(t)$ and $h_{-}(t)$, which contain the emotional expressions with positive or negative valence, respectively.
The dynamics for each component is given by: 
\begin{equation}
  \label{eq:2}
\frac{d h_{\pm}}{dt}=-\gamma_{\pm} h_{\pm}(t) + s N_{\pm}(t) + I_{\pm}(t)  
\end{equation}
where $h_{\pm}$ represents \emph{either} $h_{+}$ \emph{or} $h_{-}$.
It is increased by all agents, $N_{\pm}(t)$, that make a positive/negative contribution $(s)$ at time $t$, but can also decay over time at a rate $\gamma_{\pm}$, which reflects the decay of attention for older information.
$I_{\pm}(t)$ captures emotional information resulting from external influences, e.g. influences from the media, but is neglected in the following. 

We further define
\begin{align}
  \label{eq:11}
  h(t)=h_{+}(t)+h_{-}(t) \;; \quad \Delta h(t)=h_{+}(t)-h_{-}(t)
\end{align}
This way, the total information field $h(t)$ accumulates the emotional expressions of all agents in a weighted manner, i.e. with some memory because of the exponential decay.
$h(t)$ can be seen as a measure of the \emph{activity} of the agents, as it directly builds on their contributions. 
On the other hand, $\Delta h(t)$ is a measure of the average emotional charge, i.e. the \emph{average valence} of the information field.

\paragraph{Dynamics of valence and arousal}

As indicated in Figure~\ref{fig:schema}, the information field further influences the agent, affecting the individual valence and arousal. 
Hence, the vertical layer represents the indirect emotional communication between agents, i.e. their coupling by means of the emotional information field.  
This requires us to determine how the individual valence $v_{i}(t)$ and arousal $a_{i}(t)$ are affected by the emotional information $h$. 
For the dynamics of both variables we have proposed stochastic equations that follow the modeling framework of \emph{Brownian agents} \citep{Schweitzer2007}.
I.e. we have a superposition of \emph{deterministic} and \emph{stochastic} influences:
\begin{align}
  \label{eq:12}
\frac{d v_{i}(t)}{dt} =& - \gamma_{v}\, v_i(t) + \mathcal{G}_v +
        A_{v}\;\xi_v(t) \nonumber \\
\frac{d a_{i}(t)}{dt} =& - \gamma_{a} \, a_i(t) + \mathcal{G}_a + 
         A_{a}\,\xi_a(t)
\end{align}
Here, $A_{v}$ and $A_{a}$ denote the strength of the stochastic influences, whereas $\xi_v(t)$ and $\xi_a(t)$ are numbers randomly chosen from a standard normal distribution.
The damping constants $\gamma_{v}$, $\gamma_{a}$ ensure that in the absence of any influences both valence and arousal of an agent relax in the course of time (silent mode).
The terms $ \mathcal{G}_v$ and $ \mathcal{G}_a$, on the other hand, are \emph{non-linear functions} to capture the influence of valence and arousal, on the one hand, and the available emotional information $h$, Eq. \eqref{eq:11}, on the other hand.
To specify them, we use the general form of a power series:
\begin{alignat}{3}
  \label{eq:13}
 \mathcal{G}_{v}[h_{\pm}(t),v_i(t)] &=& \, h_{\pm}(t) \sum_{k=0}^n b_k v_{i}^{k}(t)  =& \, h_{\pm}(t)\left\{b_{0}+b_{1}v_{i}(t)+b_{2}v_{i}^{2}(t)+b_{3}v_{i}^{3}(t) \right\} \nonumber \\
      \mathcal{G}_{a}[h(t),a_i(t)] &=& h(t) \sum_{k=0}^n d_k a_{i}^{k}(t) =&  \,    h(t)\left\{d_{0}+d_{1}a_{i}(t)+d_{2}a_{i}^{2}(t) \right\}
    \end{alignat}
While this approach sounds rather abstract, it is in fact a convenient way to match the proposed dynamics with \emph{experiments}, which was successfully done in \citep{Schweitzer2016}.
    
Regarding the sign and value of the coefficients $b_{k}$, $d_{k}$, we just summarize the detailed discussion in \citep{Schweitzer2010a}.
For the dynamics of valence, we have considered contributions up to 3rd order.
If we want to ensure a ``silent mode'', $b_{0}=0$ has to be chosen.
Further, $b_{2}=0$ if we want to avoid a built-in preference for either positive or negative valence values.
With this, we have for the dynamics of valence  \citep{Schweitzer2010a}:
\begin{align}
  \label{eq:14}
\frac{d v_{i}(t)}{dt} =& \left[b_{1}h_{\pm}- \gamma_{v}\right] v_i(t) + b_{3}h_{\pm}v_{i}^{3}(t)+ 
        A_{v}\;\xi_v(t) 
\end{align}
To prevent a valence explosion, in this dynamics $b_{3}<0$, i.e. a saturation behavior for large valence values,  has to be chosen.
With this, we further see that non-trivial solutions $v\neq 0$ can be only obtained if $b_{1}>0$.
Specifically, $b_{1}>\gamma_{v}/h_{\pm}$ has to be reached. 
This condition is likely met for \emph{large} values of the emotional information $h_{\pm}$.
I.e., already from this condition we can expect a transition toward a strong emotional regime (characterized by large absolute values of valence) if the emotional information is sufficiently large.

For the dynamics of arousal, we have considered contributions up to 2nd order, i.e.:
\begin{align}
  \label{eq:15}
  \frac{d a_{i}(t)}{dt} =& \left[d_{1}h(t)- \gamma_{a} \right] a_i(t) + h(t)\left\{d_{0}+d_{2}a_{i}^{2}(t)\right\} +
         A_{a}\,\xi_a(t)
\end{align}
Our model requires a small initial positive bias, $d_{0}>0$, in order to start the communication process in the absence of previous interactions.
To allow for solutions characterized by at least two different activity levels (low and high arousal), we further have to choose $d_{1}\neq 0$ and $d_{2}\neq 0$.
The expected collective dynamics then very much depends on the \emph{sign} of $d_{2}$.
For $d_{2}<0$ the arousal dynamics becomes saturated.
If this saturation level is above the individual threshold $\tau_{i}$, the agent will generate an emotional expression and $a_{i}$ is set back to zero afterwards. 
If, at that point, fluctuations push the agent's arousal to negative values, it will not return to positive arousal again.
Hence, we obtain a scenario where agents express their emotions most likely only once.
This may lead to collective emotions, but not repeatedly.

For $d_{2}>0$, however, we may obtain two different stationary solutions with negative arousal.
At low levels of emotional information $h(t)$, e.g. after some silent periods, fluctuations are able to push the agent's arousal to positive values, from where a new communication cycle starts.
Hence, we obtain a scenario in which waves of collective emotions over time can be expected.

We note that in both cases, fluctuations play an important role in establishing an active regime.
They first push agents to a positive arousal which is then amplified by the positive feedback, until it reaches the threshold.
This then generates emotional expressions that establish a communication field which in turn feeds back on the agent's valence and arousal.
In our model \emph{valence} decides about the ``content'', determining the sign of the emotional information generated.
\emph{Arousal}, on the other hand, decides about the activity pattern.

\subsection{Modeling opinion dynamics}
\label{sec:quantifying-opinions}

We now have to specify how the emotional interaction described above influences the dynamics of \emph{opinons}.
We will proceed in two steps: first we introduce the \emph{dynamics} of opinions and subsequently the \emph{coupling} between opinions and emotions.

Specifically, our model shall explain the \emph{polarization} of opinions. 
In politics, such a polarized state can be illustrated by the opposition between \emph{democrats} ($D$) and \emph{republicans} ($R$), which each represent  particular opinions towards a given set of subjects. 
Obviously, in this context the opinion ${\theta}_{i}(t)$ of a particular agent $i$ at time $t$ cannot be represented as a binary variable, e.g. $\theta_{i}\in\{-1,+1\}$,  as this would imply that a disagreement with $R$ automatically translates into an agreement with $D$. 
Also, to extend the opinion space to include a neutral opinion, e.g. $\theta_{i}\in\{-1,0,+1\}$, does not really solves the problem, as it neglects the \emph{heterogeneity} of agents with respect to their opinions.
Therefore, a continuous variable $\theta_{i}\in (-1,+1)$ would be the most appropriate approach.

Given that opinions are continuous variables, \emph{consensus} has to be defined as a certain (narrow) \emph{range} of opinions or, more precisely, as a certain distribution  of opinions, $P(\bar{\theta},\sigma^{2})$, with a mean value $\bar{\theta}$ and a (small) variance $\sigma^{2}$.
\emph{Polarization}, on the other hand, should be defined by a \emph{bimodal} distribution of opinions $P(\theta)$, with a large variance $\sigma^{2}$. 
The two distant peaks represent the polarized opinions, while modest opinions in the middle range are less frequent.

To specify the \emph{dynamics} of opinions, we start with the general ansatz of a power series already employed in Eq. \eqref{eq:13}. 
\begin{equation}
  \label{eq:1}
  \frac{d \theta_{i}}{dt} = \sum_{k=0}^{n} \alpha_{ki} \theta_{i}^{k}(t) + A_{\theta}\xi_{\theta}(t)
\end{equation}
which postulates a non-linear feedback of the current opinion of an agent on the \emph{change} of this opinion.
The power series accounts for the fact that we need additional assumptions that later can be encoded in the coefficients $\alpha_{ki}$. 
$A_{\theta}\xi_{\theta}(t)$ is a stochastic term, to consider random influences on the opinion. 

For our discussion, we will use terms up to 3rd order from the power series. 
Neglecting individual differences and stochastic influences for the moment, we can express the opinion dynamics as: 
\begin{equation}
  \label{eq:4}
  \frac{d \theta(t)}{dt}=\alpha_{0}+\alpha_{1}\theta(t)+\alpha_{2}\theta^{2}(t)+\alpha_{3}\theta^{3}(t)
\end{equation}
To discuss the possible stationary solutions, $d\theta/dt$ of Eq. \eqref{eq:4}, for the moment we assume that $\alpha_{0}\to 0$ and $\alpha_{2}\to 0$. 
Then we find: 
\begin{equation}
  \label{eq:6}
  \hat{\theta}^{(1)}=0\;;\quad \hat{\theta}^{(2,3)}=\pm \sqrt{-\frac{\alpha_{1}}{\alpha_{3}}}
\end{equation}
\begin{figure}[htbp]
  \centering
  \includegraphics[width=0.49\textwidth]{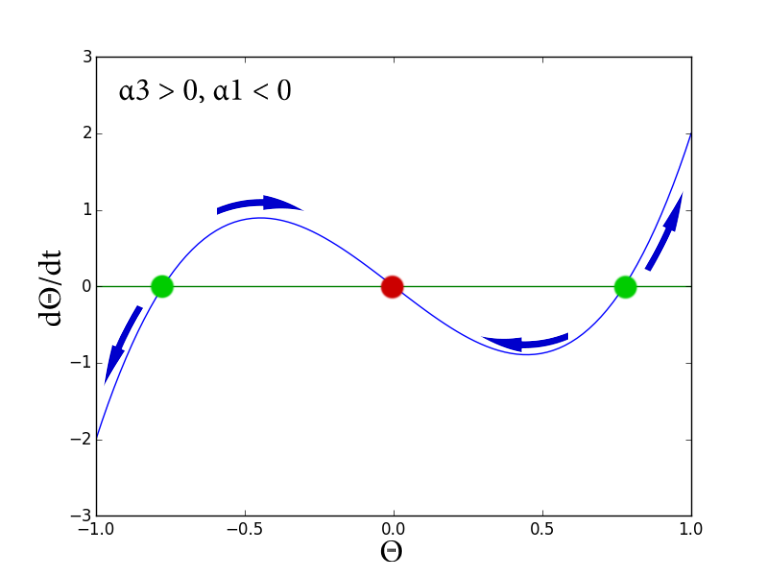}\hfill
    \includegraphics[width=0.45\textwidth]{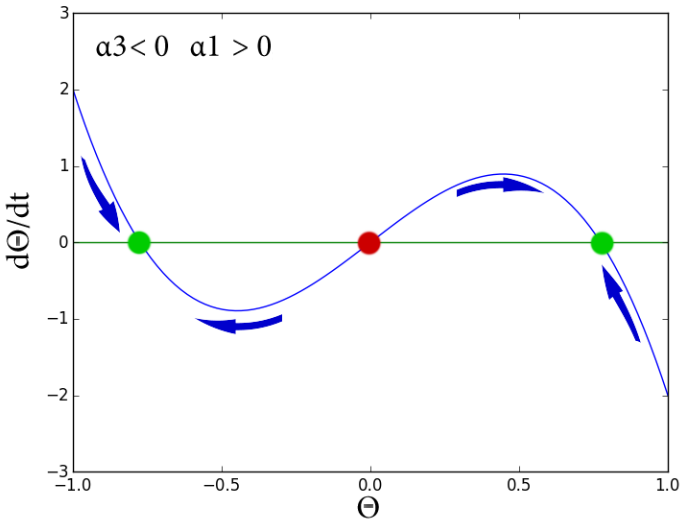}
    \caption{Plot of Eq. \eqref{eq:4} with $a_{0}=\alpha_{2}=0$ for two different parameter constellation for $\alpha_{1}$, $\alpha_{3}$. Stationary solutions for the opinions are given by colored dots. The neutral opinion (red) is an instable solution for $\alpha_{3}<0$, $\alpha_{1}>0$ and a stable solution in the opposite case. For the extreme opinions (green) the inverse stability holds.}
  \label{fig:velocity}
\end{figure}

I.e. to obtain non-trivial values for the opinion, $\alpha_{1}$ and $\alpha_{3}$ have to be of opposite sign.
The two possible cases are illustrated in Fig. \ref{fig:velocity}, where $d\theta/dt$ is plotted against $\theta$ according to Eq. \eqref{eq:4}.
We note that for $\alpha_{3}>0$, $\alpha_{1}<0$ solutions with extreme opinions are always instable.
In particular, there is a force toward the neutral opinion $\theta=0$, i.e. it would be difficult to explain \emph{polarization} of opinions based on such a parameter constellation.

\begin{figure}[htbp]
  \centering
  \includegraphics[width=0.45\textwidth]{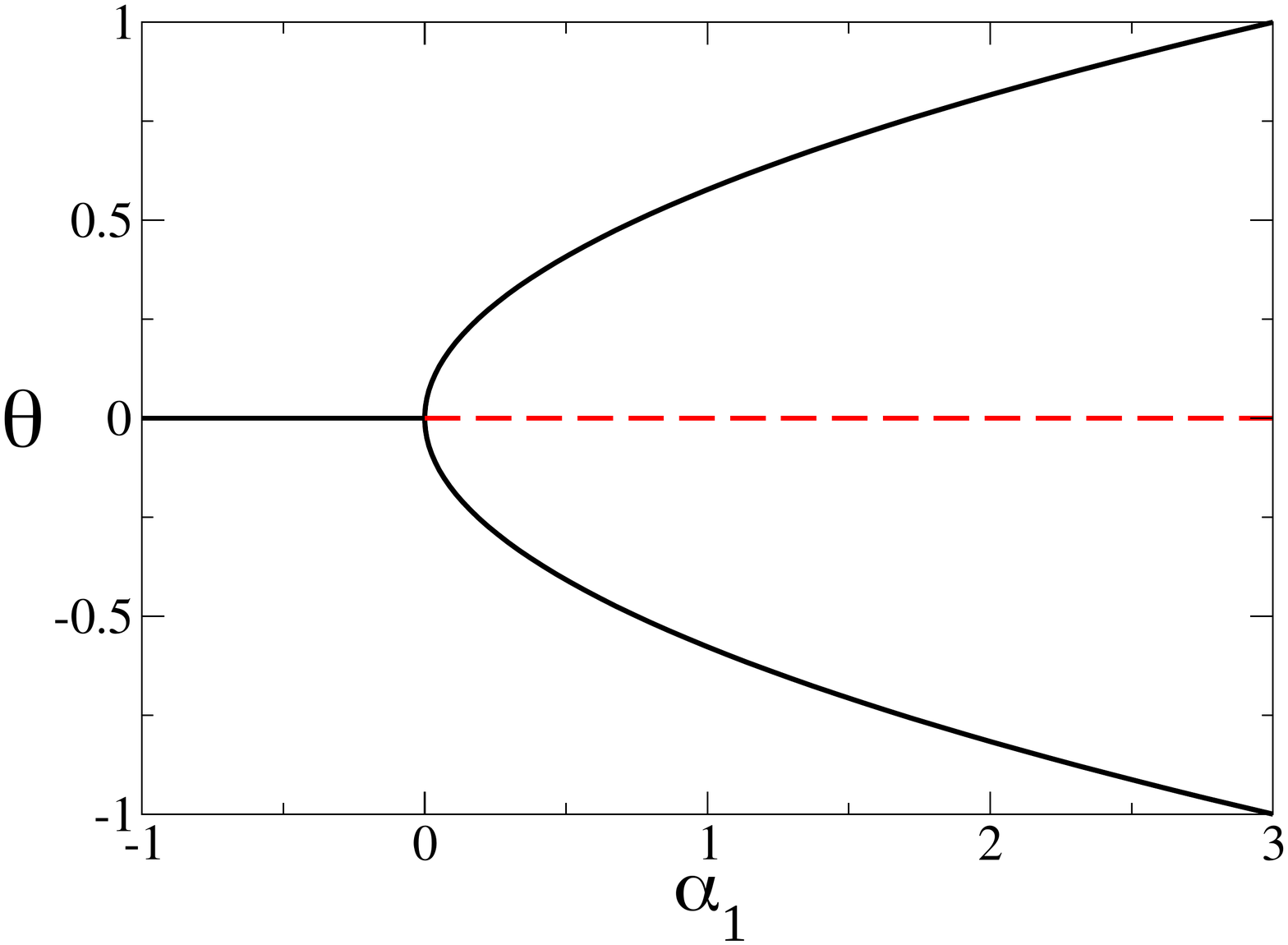}(a)\hfill
    \includegraphics[width=0.45\textwidth]{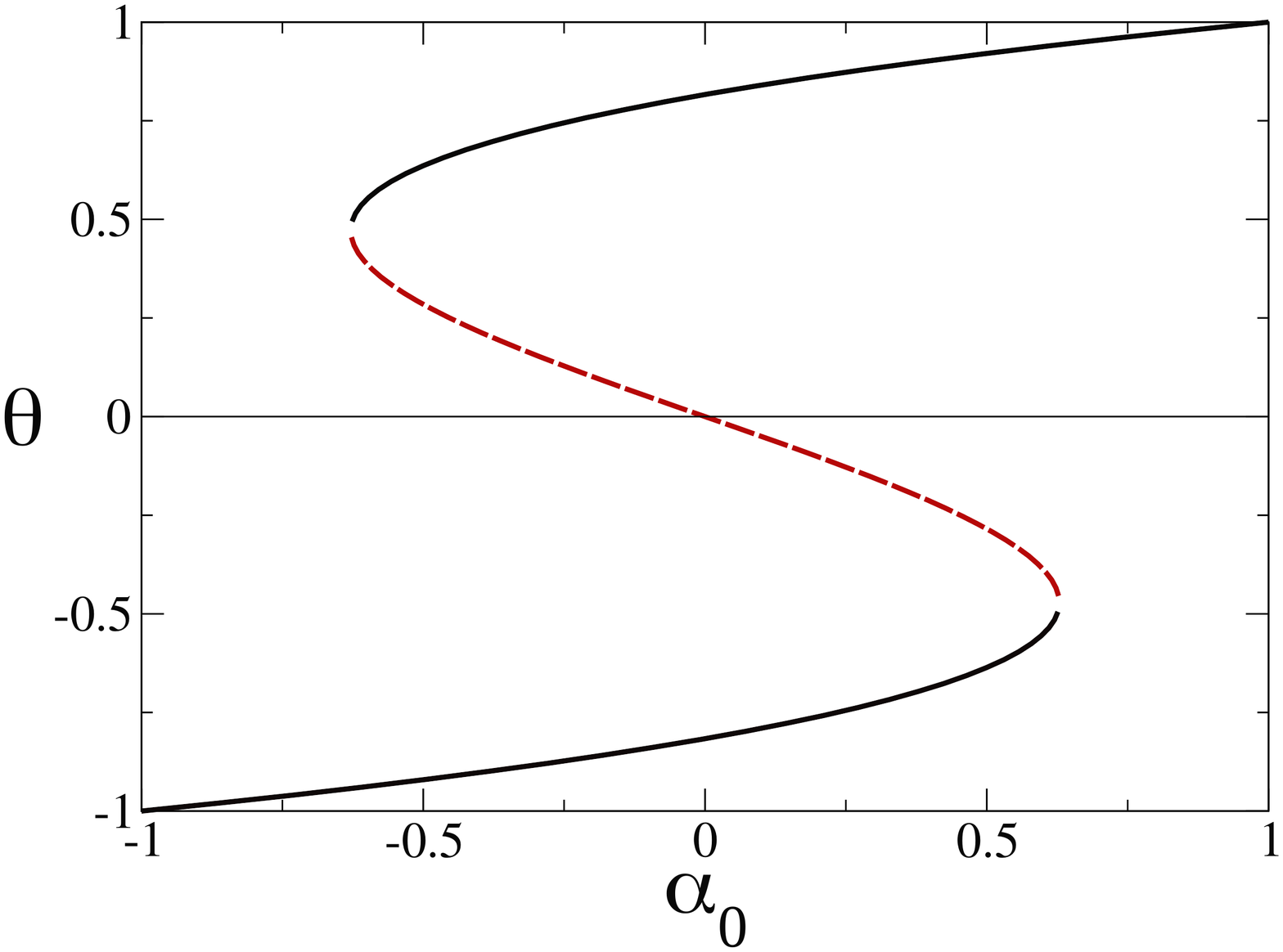}(b) 
    \caption{Bifurcation diagrams: (a) $\alpha_{0}=\alpha_{2}=0$. For $\alpha_{1}>0$, the neutral solution becomes unstable and two new stable solutions appear. (b) $\alpha_{1}=2$, $\alpha_{2}=0$. For small values of $\abs{\alpha_{0}}$, two stable solutions coexist.}
  \label{fig:bifurk}
\end{figure}

For $\alpha_{3}<0$, $\alpha_{1}>0$, on the other hand we find that there can be a stable coexistence of extreme opinions. 
Hence, in the following we choose $\alpha_{3}<0$.
This also accounts for a saturation dynamics in the case of extreme opinons.
With that, it is obvious that we will have only one possible solution for the opinion, i.e. $\hat{\theta}^{(1)}=0$, if $\alpha_{1}<0$.
However, if $\alpha_{1}>0$, we will find \textbf{bimodality}, i.e. always two opposing opinions, whereas the neutral opinion $\hat{\theta}^{(1)}=0$ is instable. 
This is the case of polarization, we are interested in.
The corresponding \emph{bifurcation diagram} is shown in Fig. \ref{fig:bifurk}(a), where we vary $a_{1}$ as the control parameter.

To understand the impact of  $a_{0}\neq 0$ on the opinion dynamics, let us now fix $\alpha_{1}>0$, i.e. we allow for polarized opinions in principle. 
For simplicity, we still neglect $a_{2}\to 0$, but we will include a non-negligible influence $a_{2}\neq 0$  later in the discussion.
The corresponding \emph{bifurcation diagram} is shown in Fig. \ref{fig:bifurk}(b), where we keep $a_{1}$ fixed, but vary $a_{0}$ as the second control parameter.
If $\alpha_{0}>0$ and $\abs{\alpha_{0}}$ is \emph{large}, then we find only one stationary solution for opinion, which is \emph{negative}, i.e. $\hat{\theta}^{(2)}=-\hat{\theta}$. 
If $\alpha_{0}<0$ and $\abs{\alpha_{0}}$ is  \emph{large}, then we find only one stationary solution for opinion, which is \emph{positive}, i.e. $\hat{\theta}^{(3)}=+\hat{\theta}$. 

For intermediate values of $\abs{\alpha_{0}}$, we find a regime where there is \textbf{bimodality} of the opinions with different weights of the positive or negative solution, and instable solutions to separate these. 
Hence, dependent on the concrete values of the $\alpha_{k}$ we can expect regimes in which we find a coexistence of polarized opinions, but as well regimes where only one opinion emerges.

\subsection{Non-linear coupling between emotions and opinions}
\label{sec:non-linear-feedback}

In order to couple the opinion dynamics to the emotional interactions of the agents, we need to determine how the coefficients $\alpha_{k}$ may depend on emotions. 
We set $\alpha_{1}(t)\propto h(t)-h_{\mathrm{base}}$, i.e. make it proportional to the overall activity in emotional interactions, expressed by the value of the emotional field.
This activity has to overcome some threshold value expressed by $h_{\mathrm{base}}$, in order to allow for a sufficient coupling.
Considering a small baseline value $h_{\mathrm{base}}$, without activity, i.e. without emotional emotional information, $\alpha_{1}$ is  negative  and there is only a neutral opinion, according to the above discussion.  
But high activity, i.e. large emotional information, will drive $\alpha_{1}$ to positive values, to allow for \textbf{bimodality}. 

In our model, $\alpha_{2}$ and $\alpha_{3}$ do not depend on the emotional dynamics.
$\alpha_{2}$ maybe close to zero. It can be positive or negative, this way generating a global bias toward left/right opinions.
$\alpha_{3}<0$ determines the saturation level of the opinions.
In agreement with Eq. \eqref{eq:6}, the smaller $\abs{\alpha_{3}}$, the larger the possible polarization of opinions.

The important parameter for the coupling between emotions and opinions is $\alpha_{0}$, for which we assume $\alpha_{0}(t)\propto -h(t) \bar{v}(t)$. 
If the emotional activity, expressed by $h(t)$ is very high and the average emotion $\bar{v}(t)$ is either very negative or very positive, then there is likely only \emph{one} opinion, which is also very likely very negative or very positive. 
Precisely, if $\bar{v}<0$ and $h$ is high, then $\alpha_{3}>0$ and $\abs{\alpha_{3}}$ is large and we find  $\hat{\theta}^{(2)}=-\hat{\theta}$. 
If $\bar{v}>0$ and $h$ is high, then $\alpha_{3}<0$ and $\abs{\alpha_{3}}$ is large and we find  $\hat{\theta}^{(3)}=\hat{\theta}$. 
In accordance with Fig. \ref{fig:bifurk}(b), for moderate values of $\bar{v}$ and $h$ we can expext \textbf{bimodality}, i..e a polarization of opinions.

In conclusion, the dynamics of the opinions read now
\begin{equation}
  \label{eq:5}
  \frac{d\theta(t)}{dt}= -c_{0}h(t)\bar{v}(t)  +c_{1}\left[h(t)-h_{\mathrm{base}}\right]\theta(t)+ \alpha_{2}\theta^{2}(t)
  + \alpha_{3} \theta^{3}(t) + A_{\theta}\xi_{\theta}(t)
\end{equation}
where $c_{0}$, $c_{1}$ are some proportionality constants.
They are chosen between 0 and 1 to mitigate the influence of the corresponding terms.

\paragraph{Relations to the bounded confidence model. \ }

To further understand this dynamics, let us neglect all small or higher-order terms and focus on the core dynamics, which reads for an individual agent $i$:
\begin{equation}
  \label{eq:5a}
  \frac{d\theta_{i}(t)}{dt}= h(t) \big[c_{1}\theta_{i}(t)- c_{0}\bar{v}(t) \big]
  \end{equation}
For the \emph{mean opinion} in the agent population, we find:
\begin{equation}
  \label{eq:8}
  \frac{d\bar{\theta}(t)}{dt}= \frac{1}{N}\sum_{i} \frac{d\theta_{i}(t)}{dt}= h(t)  \left[\frac{1}{N}\sum_{i}c_{1}\theta_{i}(t)-c_{0}\bar{v}(t)\right]
\end{equation}
This leads, in \emph{equilibrium}, ${d\bar{\theta}/dt}=0$, to $\bar{\theta}=(c_{0}/c_{1})\bar{v}$, and 
the opinion dynamics, Eq. \eqref{eq:5a}, simplifies to:
\begin{equation}
  \label{eq:9}
  \frac{d\theta_{i}(t)}{dt}= \mu \big[\theta_{i}(t) - \bar{\theta}(t)\big]
\end{equation}
which is precisely the bounded confidence dynamics with $\mu=c_{1}h(t)$. 
That means, the higher the activity as measured by $h(t)$, the faster the convergence toward the mean. 
A noticeable difference: in the bounded confidence model, agents only interact if $\abs{\theta_{i}-\theta_{j}}<\epsilon$.
In our case, all agents interact through the communication field $h(t)$, more precisely via the mean valence $\bar{v}(t)$.
In the following, we make use of the fact that the average valence $\bar{v}$ can be approximated by the difference in the two field components, $h_{+}$, $h_{-}$, Eq. \eqref{eq:11} and choose
\begin{align}
  \label{eq:17}
  \bar{v}(t)=c_{0}\Delta h(t)
\end{align}
which reduces the number of variables and parameters. 

Very similar to the bounded confidence model, we should expect scenarios that can lead to \emph{consensus} characterized by a \emph{unimodal} opinion distribution.
However, as we already know from the above discussion, the \emph{inclusion} of the terms $\alpha_{2}$, $\alpha_{3}$ neglected in this derivation may lead to other scenarios of \emph{coexistence} characterized by a bimodal opinion distribution.
Hence, while $\alpha_{0}$ and $\alpha_{1}$ are essential to couple the opinion dynamics to the emotion dynamics, $\alpha_{2}$ and $\alpha_{3}$ may decide about the level of polarization reached.

\section{Results}
\label{sec:results-comp-simul}

\subsection{Agent-based simulations}
\label{sec:agent-based-simul}

We first present the results of agent-based computer simulations, to verify that the model works as expected.
The parameter values not explicitly mentioned in the following are chosen as follows:
\begin{align}
\label{parameters-new}
\begin{split}
\gamma_v = 0.5,  b_1=1, b_3=-1, A_v=0.3, \xi_{v}\sim \mathcal{N}(0, 0.5)\\
\gamma_a = 0.9,  d_0=0.05, d_1=0.5, d_2=0.1, A_a=0.3,  \xi_{a}\sim \mathcal{N}(0, 6)\\ \gamma_h = 0.7, s=0.6, h_{\mathrm{base}} = 0.1 \\ \tau_{min} = 0.1, \tau_{max} =1.1, N=100, \delta t=0.2 \\
 c_{0}=0.1, c_{1}=1, A_{\theta} = 0.05
\end{split}
\end{align}

\begin{figure}[htbp]
  \centering
  \includegraphics[width=0.45\textwidth]{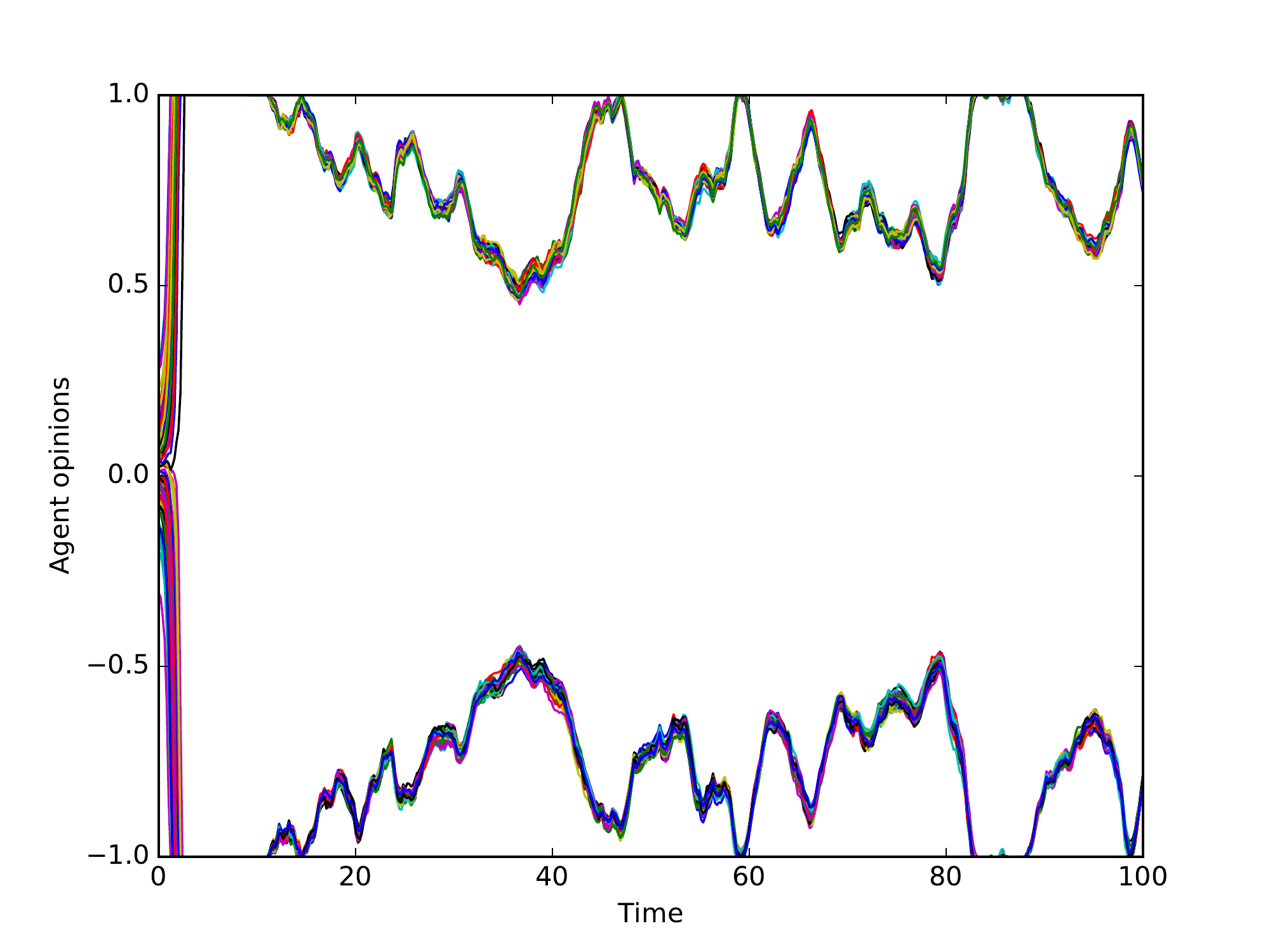}(a) \hfill 
    \includegraphics[width=0.45\textwidth]{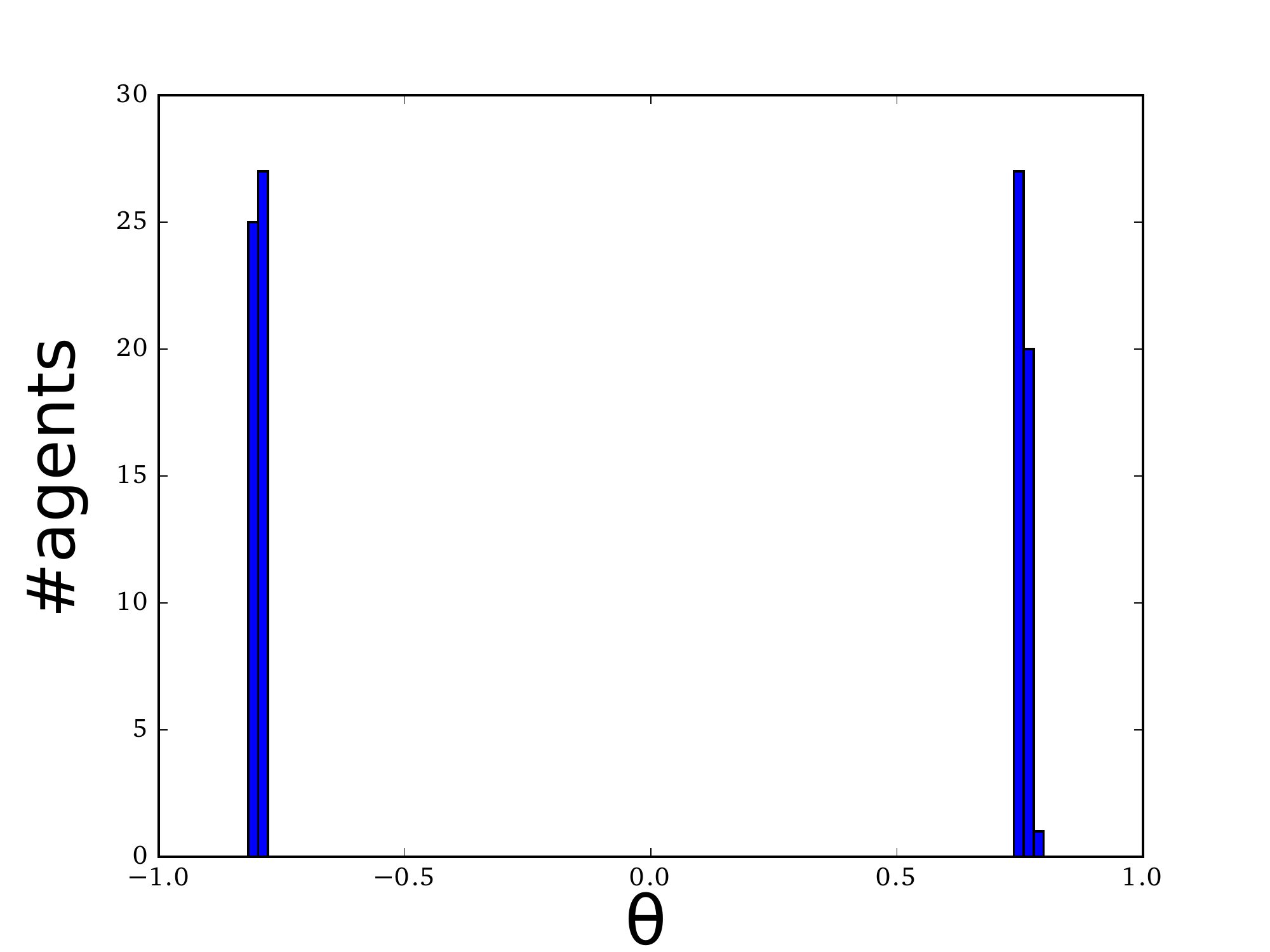}(c) \\
  \includegraphics[width=0.45\textwidth]{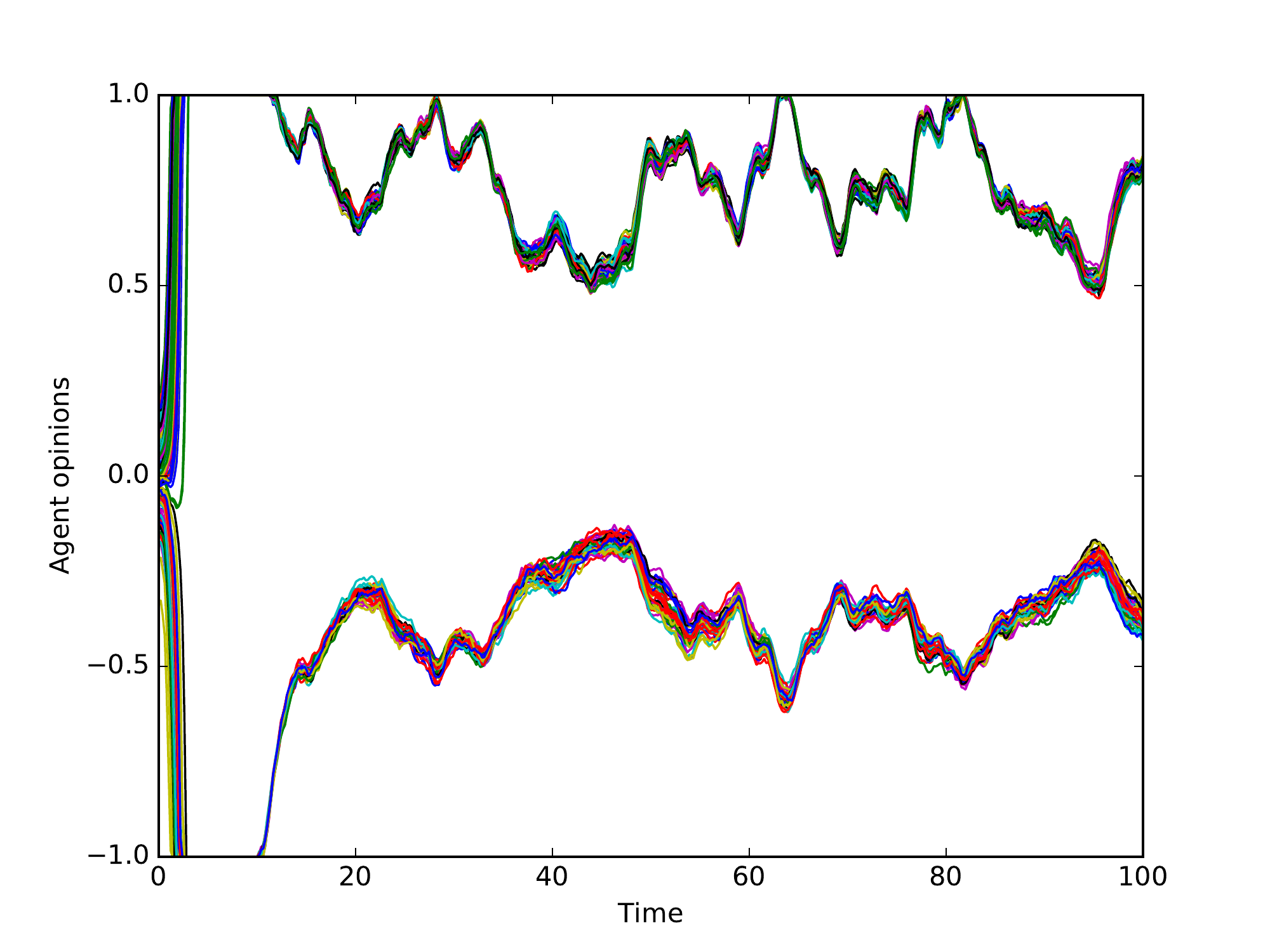}(b) \hfill 
    \includegraphics[width=0.45\textwidth]{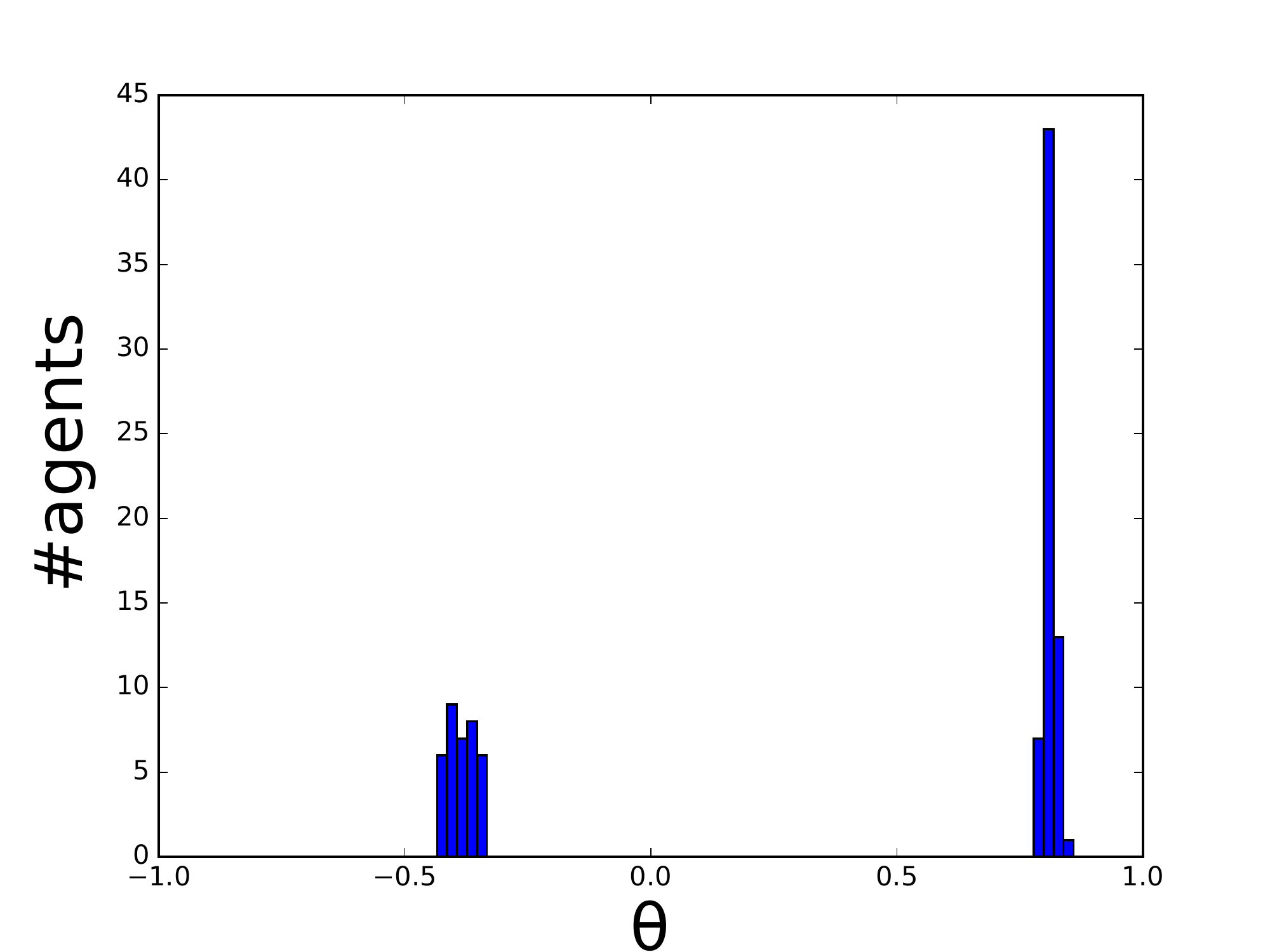}(d)
  \caption{(a,b) Opinion trajectories, (c,d) Opinion distribution at $t=100$. (a,c) $\alpha_{2}=0$, (b,d) $\alpha_{2}=2$, other parameters: $\alpha_{3}=-5$.}
  \label{fig:simul}
\end{figure}

Figure \ref{fig:simul} shows, in the left panel, the opinion trajectories of individual agents over time and, in the right panel, the distribution of opinions after a sufficiently long time.
We note that, because the model is intrinsically stochastic, we do not reach a stationary distribution in the strict sense.

As the first observation, we find indeed the \emph{polarization of opinions} that were initially close, i.e. drawn from a normal distribution $\mathcal{N}(\mu,\sigma^{2})=\mathcal{N}(0,0.3)$.
This was expected because the range of parameters was adjusted such that a polarization regime emerges.
However, we emphasize that some of the $\alpha_{k}$, namely $\alpha_{0}$ and $\alpha_{1}$, are in fact not constants, but functions of the emotional field components $h_{+}$, $h_{-}$ which in turn depend on the agents' individual valence and arousal.
Hence, these ``parameters'' were not chosen, but their value emerged from the emotional interactions between agents.
Moreover, these values are not fixed but fluctuate over time according to the emotional dynamics.
This leads to the non-stationary opinion dynamics observed.

As the second observation, we see that the number of agents with positive and negative opinions can differ significantly dependent on the parameter $\alpha_{2}$. Although it is small, it generates a global preference for either left or right opinions. 
Consequently, we find also the emergence of a \emph{minority/majority} in the agent population.
This is shown in the different heights of the peaks of the bimodal distributions.

\begin{figure}[htbp]
  \centering
  \includegraphics[width=0.45\textwidth]{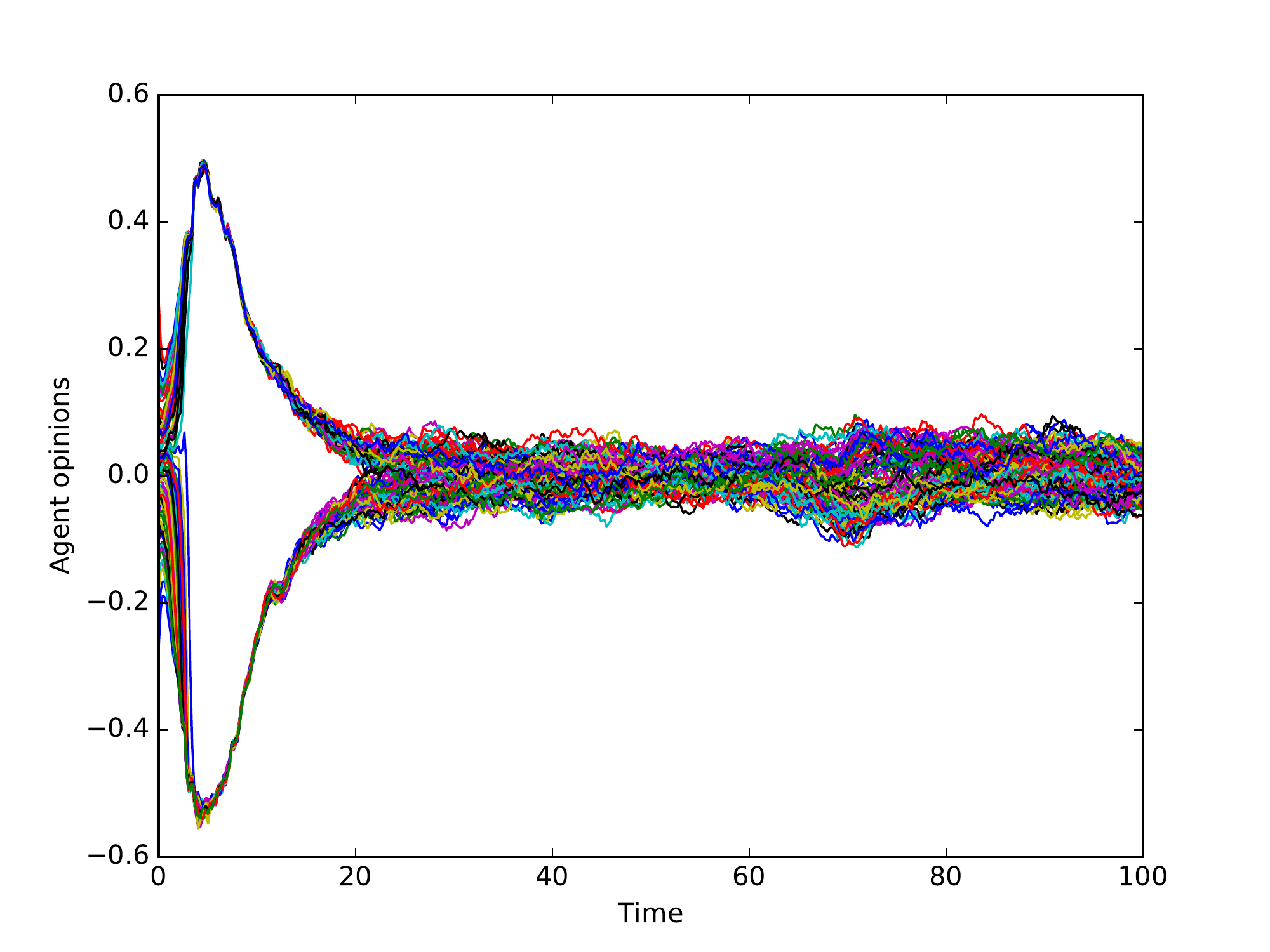}(a) \hfill
  \includegraphics[width=0.45\textwidth]{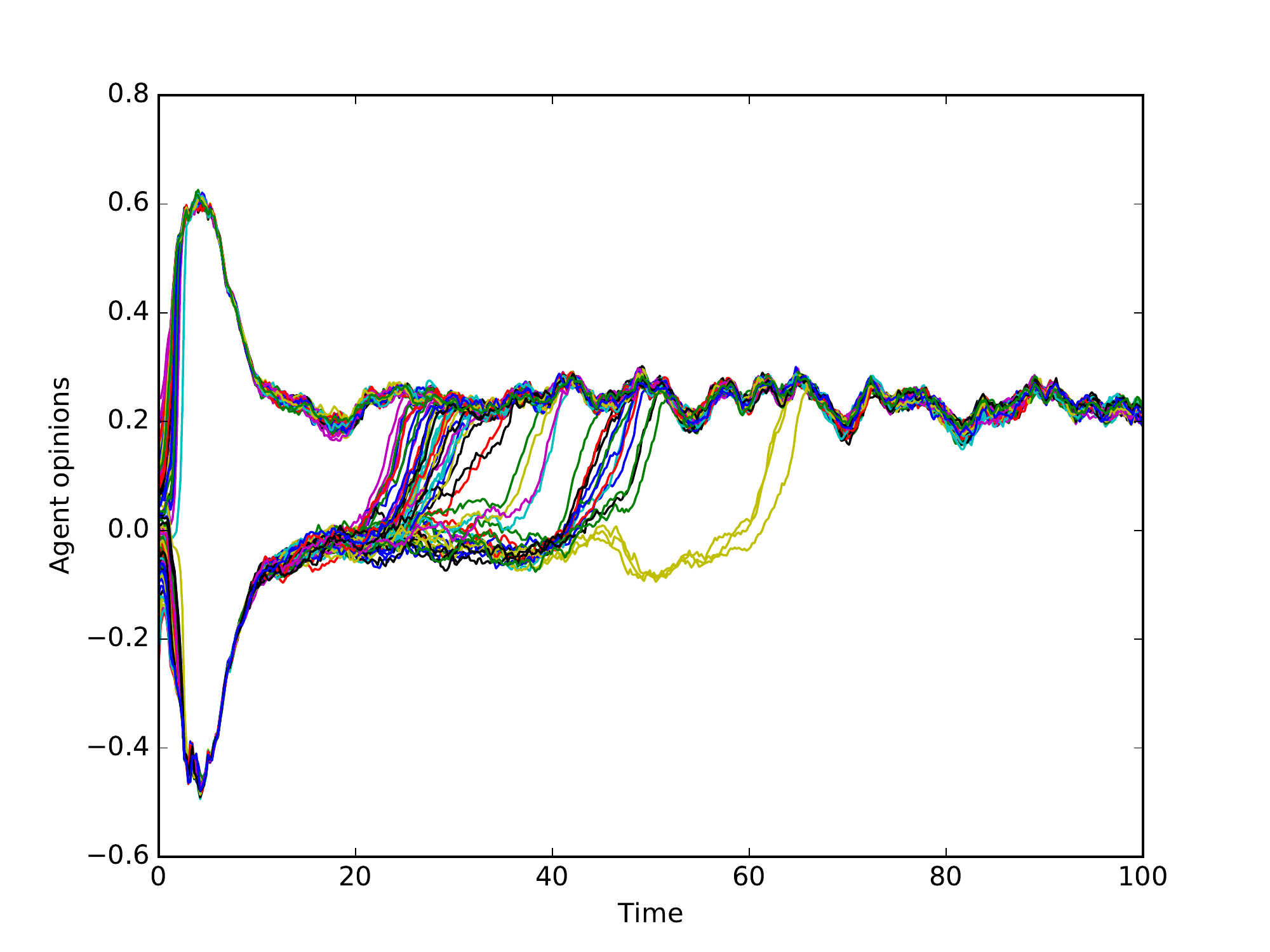}(b)
  \caption{Opinion trajectories for the case of \emph{consensus}. (a) $\alpha_{2}=0$, hence the neutral opinion $\theta=0$ is obtained. (b) $\alpha_{2}=4$, hence a global bias toward positive opinions exists.}
  \label{fig:consesus}
\end{figure}

Eventually, we can also obtain scenarios in which consensus is reached, i.e. instead of a bimodal opinion distribution we find a unimodal distribution.
This is illustrated in Figure \ref{fig:consesus} for a consensus around the neutral opinion and and a biased opinion.

\subsection{Critical transitions toward polarized opinions}
\label{sec:crit-trans-toward}

\begin{figure}[h!]
  \centering
  \includegraphics[width=0.8\linewidth]{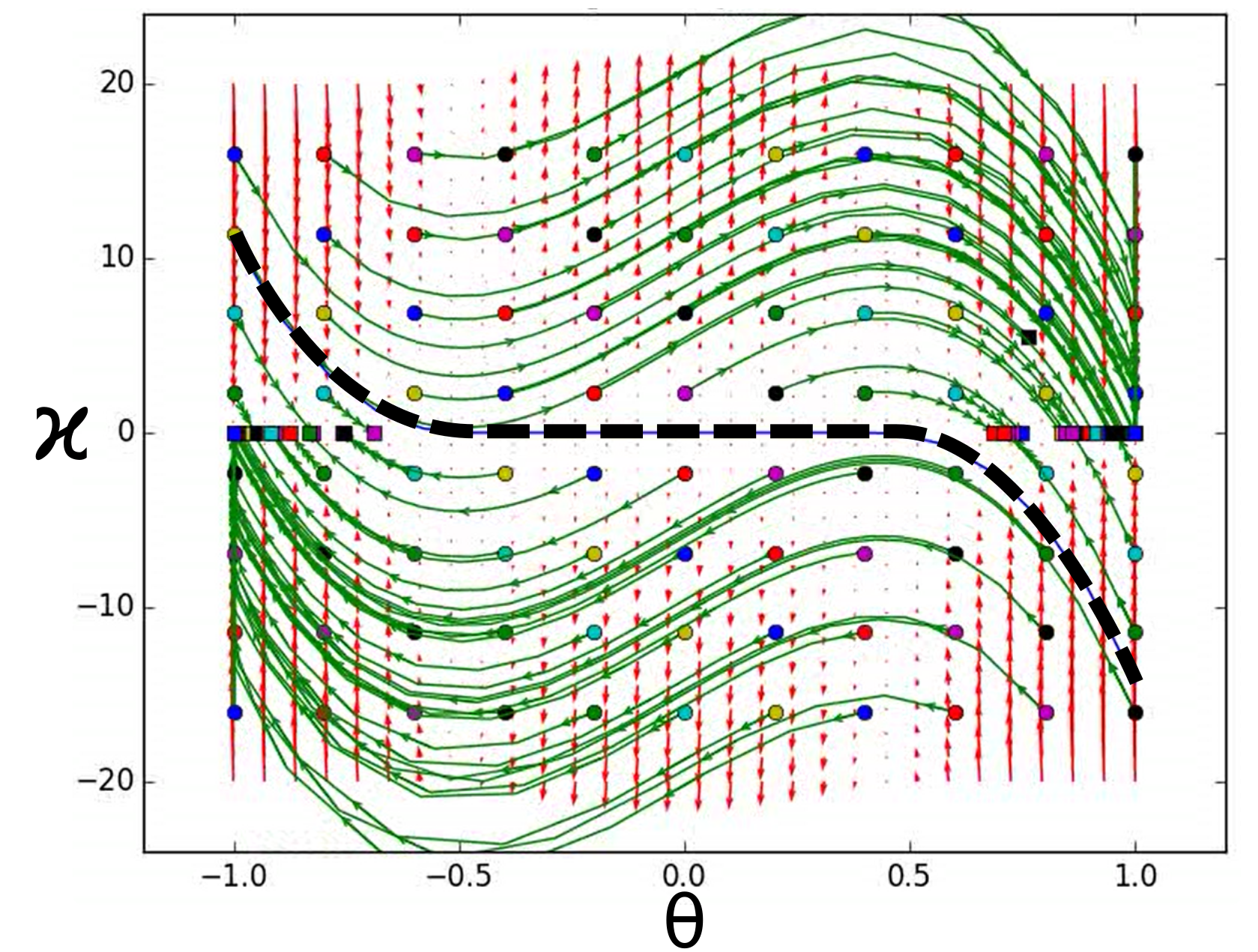}
  \caption{Phase portrait $\big(\kappa(\theta),\theta\big)$ obtained from solving the coupled Eqs. \eqref{omega_dot} 
    in the interval $\theta\in[-1,1]$. Parameters $a_{0}=0$, $\alpha_{1}=15$, $\alpha_{2}=-2$, $\alpha_{3}=-23$. Solutions (green lines) run from numerous starting points (colored circles) to end points (colored squares).
    The black dashed line  depicts the separatrix given by Eq. \eqref{separatrix}.
    The values of $\theta_{\pm}$, Eq. \eqref{critical2}, are visible by the  vertical regions in which $d\kappa/dt=0$ (i.e, no arrows visible). }
  \label{fig:typical}
\end{figure}

To fully understand the role of the coefficients $\alpha_{k}$ in Eq. \eqref{eq:4}, we now focus on the so-called \emph{phase portrait}.
Different from Fig. \ref{fig:velocity}, which plots $d\theta/dt$ against $\theta$, the phase portrait investigates $d^{2}\theta/dt^{2}$ against $d\theta/dt$.
With Eq. \eqref{eq:4}, the two variables of the phase portrait follow the dynamics:
\begin{align}
  \frac{d\theta(t)}{dt}&=\kappa(\theta)=\alpha_{0}+\alpha_1\theta(t) +\alpha_2 \theta^{2}(t) +\alpha_3 \theta^3(t)
  \nonumber \\
  \label{omega_dot}
  \frac{d\kappa(\theta)}{dt}&=\kappa(\theta)\, \big[\alpha_1 +2\alpha_2 \theta(t) +3 \alpha_3 \theta^2(t)\big]
\end{align}
These two coupled equations can be solved numerically using a 4th order Runge-Kutta method.
The results are shown in Fig. \ref{fig:typical}.
Stationary solutions of the opinions are given by $\kappa(\theta)=0$.
But, as the distribution of the colored squares along this horizontal line in Fig. \ref{fig:typical} verifies, the solutions concentrate on the left/right end of the horizontal line, indicating polarized opinions.
The middle-ranged values of $\theta$ resulting from $\kappa(\theta)=0$ are in fact unstable stationary solutions. 

The range of these unstable solutions can be obtained by setting $\big[\alpha_1 +2\alpha_2 \theta(t) +3 \alpha_3 \theta^2(t)\big]=0$ in Eq. \eqref{omega_dot}. 
As the result we find:
\begin{align}
\kappa=0\;\; \text{ is } \;\; \left\{
\begin{array}{clcc}
\text{stable} & \text{if} & \theta <  \theta_-\\
\text{instable} & \text{if} &\theta_- < \theta<\theta_+\\
\text{stable} & \text{if} & \theta_+ < \theta\\
\end{array}
  \right.
\end{align}
where $\theta_{+}$ and $\theta_{-}$ follow from the above quadratic equation: 
\begin{align}\label{critical2}
  \theta_{\pm}=-\frac{\alpha_2}{3\alpha_3}\pm \frac{1}{3\alpha_3}D \;; \quad D=\sqrt{\alpha_2^2-3\alpha_1\alpha_3}
\end{align}
These values are clearly indicated in Fig. \ref{fig:typical} by the two (empty) \emph{vertical regions} around $\theta_{\pm}\approx \pm 0.5$ in which the arrows of the vector field change their direction.

So, where do agents end up with their opinions dependent on their initial conditions, if we only consider the deterministic dynamics?
This is answered by the \emph{separatrix} also depicted in Fig. \ref{fig:typical} as the thick dashed line. 
Agents starting with initial conditions above the separatrix will tend to obtain a positive $\theta$ in the long run, while agents starting with initial conditions below the separatrix will reach  a negative $\theta$.
The precise equation for the separatrix is given by
\begin{equation}\label{separatrix}
  S(\theta)=
  \left\{
\begin{array}{clcc}
    \alpha_0^-+\alpha_1 \theta +\alpha_2  \theta^2 +\alpha_3  \theta^3       &  \text{if } & \theta<\theta_-\\
    0 &  \text{if } & \theta_-<\theta<\theta_+\\
    \alpha_0^++\alpha_1  \theta +\alpha_2  \theta^2 +\alpha_3  \theta^3       &  \text{if } & \theta_+<\theta,\\
\end{array}
\right.
\end{equation}
The only difference in the expression is in the values $\alpha_0^-$, $\alpha_0^+$, which, when taken at first order in $\alpha_2$, can be reduced to
\begin{align}
  \label{alpha_pm}
  \alpha_0^\pm =
  \frac{-2\alpha_2^3+9\alpha_1\alpha_2\alpha_3\pm2 \alpha_2^2 D \mp 6\alpha_1\alpha_3 D}{27\alpha_3^2}
  \,\approx\, \frac{3 \alpha_1\alpha_2 \mp 2 \alpha_1 D}{9 \alpha_3}.
\end{align}
It is important to notice that the dynamics captured by the phase portrait, shown in Fig. \ref{fig:typical}, does not depend on $\alpha_0$. The coefficient $\alpha_0$ merely selects a solution curve and can be seen as a vertical shift in the diagram. 

\section{Discussion}
\label{sec:discussion}

In this paper, we have provided a model to formally link the dynamics of \emph{emotions} to the dynamics of \emph{opinions}.
We follow an \emph{agent-based approach}, that is, we focus on the emotions and opinions of individual agents with \emph{heterogeneous} properties.
Our research interest is to explain the emergence of \emph{collective opinions} based on the \emph{emotional interaction} between agents.
More specifically, we want to understand under which conditions we obtain \emph{consensus}, i.e. the emergence of one common opinion (reflected in a narrow opinion distribution), or \emph{polarization}, i.e. the emergence of two opposing common opinions (reflected in a bimodal opinion distribution).
As the interaction mechanism, we do \emph{not} assume a feedback between different \emph{opinions}, which would be the most simple way to obtain the two distinct opinion distributions.
Instead, our main assumption is that the dynamics of \emph{opinions} is driven by the dynamics of \emph{emotions}.

Following established measures from social psychology, the dynamics of emotions is characterized by two agent variables, \emph{valence}, the pleasure associated with emotions, and \emph{arousal}, the activity associated with emotions.
Both variables determine the emotional expression of agents, from which \emph{collective emotional information} is generated.
This is quantified by two aggregated and time dependent variables, the emotional field $h(t)$ and the average valence, $\bar{v}(t)$.
These two variables from the emotional interaction feed back on the individual opinion dynamics in a non-linear manner.
Our formal model makes transparent under which critical conditions for the \emph{emotional interaction} we can expect a \emph{polarization of opinions}, without assuming a direct interaction on the level of opinions.

Our modeling approach fits a general framework to model \emph{active matter} \citep{Schweitzer2018}, a term to denote systems with the ability of  self-organization, active motion and structure formation, provided a critical supply of energy.
In our case, the \emph{driving} variables describe the emotional state, $\{a_{i}(t),v_{i}(t)\}$ composed of valence and arousal, whereas the \emph{driven} variable is the individual opinion,  $\theta_{i}(t)$.
Different from other approaches, our model respects the fact that emotions and opinions evolve on two different time scales.
Emotions relax faster than opinions, i.e. they evolve on a shorter time scale, and the resulting values for the mean valence and the emotional field subsequently drive the evolution of opinions.

\paragraph{Validation scenario}

Our investigations point to real world mechanisms in the formation of opinions which, in addition to rational considerations, are very much dependent on sentiment.
Because the measurement of emotions is simpler than a corresponding measurement of opinions (see Sect. \ref{sec:Background}), our model can be potentially validated against data.
For instance, we are able to measure the emotional information from \emph{online communication} in fora.
From the user comments in fora valence can be quantified by means of refined sentiment analysis tools, such as lexicon-based techniques.
This would allow to estimate $\bar{v}$ as a real value between $-1$ and $1$.
To estimate the value of the emotional information $h(t)$, we can use the  total amount, or even a time series, of comments.
This would also allow to estimate values for the parameter $\gamma$ for the relaxation  dynamics. 

In order to validate the link between emotion dynamics and opinion dynamics, we also need to estimate the \emph{polarization of opinions}.
Here, we see   three alternatives to quantify polarization:
\begin{description}
\item[\rm \em Polarization from the final amount of likes/dislikes:]
  This dichotomoy constrains opinions as strictly positive or negative, not neutral.
  Data, e.g. from  \texttt{Reddit} or \texttt{Youtube} allows us to calculate the ratio of agents with positive and negative opinions and to link this ratio to different collective emotion scenarios (expressed by $h$ and $\bar{v}$). 

\item[\rm \em Polarization from the coexistence of positive and negative expressions  of opinions:]
  Using a signed lexicon, we can
  estimate the opinion of an individual towards a topic on a real scale. 
  For very big and longitudinal data, we can obtain the final distribution of opinions $P(\theta)$, to calculate the overall polarization.
  Changes in the distribution and the subsequent polarization measure could then be related to changes in the emotional information.
  
\item[\rm \em Polarization from the existence of network components:]
  The online communication between users can be represented as a social network, on which we can perform a community analysis, to detect communities with different opinions.
This method makes sense if links signal agreement or endorsement, like retweets or follower links.
  \end{description}

\paragraph{Extension to multi-dimensional opinion space}

Our model so far considers that opinion is a \emph{scalar}, i.e. there is only \emph{one} opinion per agent, which is with respect to one subject only.
This one-dimensional description already grafts the dichotomy \emph{in favor/against} into the opinion space. 
In most real situations, however, one could agree with individual $i$ on one particular subject and with $j$ on another one. 
Hence, more complex \emph{mixed opinions} should be possible in an extension of our model. 
For this we could redefine the opinion of agent $i$ as a \emph{vector} $\mathbf{\theta}_{i}(t)=\{\theta_{i|1}(t),\theta_{i|2}(t),...\}$, where $\theta_{i|n}(t)$ expresses the opinion of agent $i$ towards subject $n$ at time $t$. 
It then depends on the context how emotions drive opinions in different dimensions. 

We emphasize that a multi-dimensional opinion space exacerbates the problem of \emph{consensus}, i.e. the \emph{convergence} toward a common opinion. 
Convergence along one dimension does not necessarily implies agreement on other subjects. 
On the contrary, agents which agree on given subjects often choose to disagree on other subjects, to distinguish themselves from other agents.
Hence, we expect that instead of consensus we frequently find \emph{coexistence} of (mixed) opinions.    
In developed democracies such as Switzerland, this has lead to the emergence of a political space with many parties coexisting, which opens the possibility to form alliances regarding certain decisions \citep{Abisheva2015}. 
 
It is an open question how to define a \emph{multidimensional polarization measure} based on mixed opinions.
This will be addressed in a subsequent publication.
Another open question regards the link of such measures to available data.
For this, we could consider a topic model to reduce the dimensionality of the opinion space.
With a smaller number of opinion dimensions, we can then test correlations to identify additional polarization levels.

\subsection*{Acknowledgments}
All authors acknowledge funding from the Swiss National Science Foundation (CR21I1\_146499).
D.G. acknowledges funding from the Vienna Science and Technology Fund through the Vienna Research Group Grant ``Emotional Well-Being in the Digital Society'' (VRG16-005).

\small \setlength{\bibsep}{1pt}


\begin{thebibliography}{46}
\expandafter\ifx\csname natexlab\endcsname\relax\def\natexlab#1{#1}\fi
\expandafter\ifx\csname url\endcsname\relax
  \def\url#1{\texttt{#1}}\fi
\expandafter\ifx\csname urlprefix\endcsname\relax\def\urlprefix{URL }\fi
\expandafter\ifx\csname selectlanguage\endcsname\relax
  \def\selectlanguage#1{\relax}\fi

\bibitem[{Ahn \emph{et~al.}(2012)Ahn, Gobron, Garcia, Silvestre, Thalmann and
  Boulic}]{Ahn2012}
Ahn, J.; Gobron, S.; Garcia, D.; Silvestre, Q.; Thalmann, D.; Boulic, R.
  (2012).
\newblock An NVC emotional model for conversational virtual humans in a 3D
  chatting environment.
\newblock In: \emph{International Conference on Articulated Motion and
  Deformable Objects}. Springer, pp. 47--57.

\bibitem[{Berger(2011)}]{Berger2011}
Berger, J. (2011).
\newblock Arousal increases social transmission of information.
\newblock \emph{Psychological Science} \textbf{22(7)}, 891--893.

\bibitem[{Bollen \emph{et~al.}(2011)Bollen, Mao and Zeng}]{Bollen2011}
Bollen, J.; Mao, H.; Zeng, X. (2011).
\newblock Twitter mood predicts the stock market.
\newblock \emph{Journal of Computational Science} \textbf{2(1)}, 1--8.

\bibitem[{Castellano \emph{et~al.}(2009)Castellano, Fortunato and
  Loreto}]{Castellano2009}
Castellano, C.; Fortunato, S.; Loreto, V. (2009).
\newblock Statistical physics of social dynamics.
\newblock \emph{Reviews of Modern Physics} \textbf{81(2)}, 591.

\bibitem[{Chmiel \emph{et~al.}(2011)Chmiel, Sienkiewicz, Thelwall, Paltoglou,
  Buckley, Kappas and Ho{\l}yst}]{Chmiel2011}
Chmiel, A.; Sienkiewicz, J.; Thelwall, M.; Paltoglou, G.; Buckley, K.; Kappas,
  A.; Ho{\l}yst, J.~A. (2011).
\newblock Collective emotions online and their influence on community life.
\newblock \emph{PloS ONE} \textbf{6(7)}, e22207.

\bibitem[{Dandekar \emph{et~al.}(2013)Dandekar, Goel and Lee}]{Dandekar2013}
Dandekar, P.; Goel, A.; Lee, D.~T. (2013).
\newblock Biased assimilation, homophily, and the dynamics of polarization.
\newblock \emph{Proceedings of the National Academy of Sciences}
  \textbf{110(15)}, 5791--5796.

\bibitem[{Fontaine \emph{et~al.}(2007)Fontaine, Scherer, Roesch and
  Ellsworth}]{Fontaine2007}
Fontaine, J.~R.; Scherer, K.~R.; Roesch, E.~B.; Ellsworth, P.~C. (2007).
\newblock The world of emotions is not two-dimensional.
\newblock \emph{Psychological Science} \textbf{18(12)}, 1050--1057.

\bibitem[{Forgas(2008)}]{Forgas2008}
Forgas, J.~P. (2008).
\newblock Affect and cognition.
\newblock \emph{Perspectives on Psychological Science} \textbf{3(2)}, 94--101.

\bibitem[{Fortunato \emph{et~al.}(2013)Fortunato, Macy and
  Redner}]{Fortunato2013}
Fortunato, S.; Macy, M.; Redner, S. (2013).
\newblock Editorial.
\newblock \emph{Journal of Statistical Physics} \textbf{151(1-2)}, 1--8.

\bibitem[{Frijda(1987)}]{Frijda1987}
Frijda, N.~H. (1987).
\newblock Emotion, cognitive structure, and action tendency.
\newblock \emph{Cognition and Emotion} \textbf{1(2)}, 115--143.

\bibitem[{Garas \emph{et~al.}(2012)Garas, Garcia, Skowron and
  Schweitzer}]{Garas2012}
Garas, A.; Garcia, D.; Skowron, M.; Schweitzer, F. (2012).
\newblock {Emotional persistence in online chatting communities}.
\newblock \emph{Scientific Reports} \textbf{2}, 402.

\bibitem[{Garcia \emph{et~al.}(2015)Garcia, Abisheva, Schweighofer, Serdult and
  Schweitzer}]{Abisheva2015}
Garcia, D.; Abisheva, A.; Schweighofer, S.; Serdult, U.; Schweitzer, F. (2015).
\newblock Ideological and Temporal Components of Network Polarization in Online
  Political Participatory Media.
\newblock \emph{Policy and Internet} \textbf{7(1)}, 46--79.

\bibitem[{Garcia \emph{et~al.}(2016)Garcia, Kappas, Kuster and
  Schweitzer}]{Schweitzer2016}
Garcia, D.; Kappas, A.; Kuster, D.; Schweitzer, F. (2016).
\newblock The Dynamics of Emotions in Online Interaction.
\newblock \emph{Royal Society Open Science} \textbf{3(160059)}.

\bibitem[{Garcia and Schweitzer(2011)}]{Garcia2011}
Garcia, D.; Schweitzer, F. (2011).
\newblock Emotions in product reviews--empirics and models.
\newblock In: \emph{Privacy, Security, Risk and Trust (PASSAT) and 2011 IEEE
  Third Inernational Conference on Social Computing (SocialCom), 2011 IEEE
  Third International Conference on}. IEEE, pp. 483--488.

\bibitem[{Garcia and Schweitzer(2015)}]{Garcia2015}
Garcia, D.; Schweitzer, F. (2015).
\newblock Social signals and algorithmic trading of Bitcoin.
\newblock \emph{Royal Society Open Science} \textbf{2(9)}, 150288.

\bibitem[{Goldenberg \emph{et~al.}(2014)Goldenberg, Saguy and
  Halperin}]{Goldenberg2014}
Goldenberg, A.; Saguy, T.; Halperin, E. (2014).
\newblock How group-based emotions are shaped by collective emotions: Evidence
  for emotional transfer and emotional burden.
\newblock \emph{Journal of Personality and Social Psychology} \textbf{107(4)},
  581.

\bibitem[{Holyst(2016)}]{holyst2016cyberemotions}
Holyst, J.~A. (2016).
\newblock \emph{Cyberemotions: Collective Emotions in Cyberspace}.
\newblock Springer.

\bibitem[{Kappas(2013)}]{Kappas2013}
Kappas, A. (2013).
\newblock Social regulation of emotion: messy layers.
\newblock \emph{Frontiers in Psychology} \textbf{4}, 51.

\bibitem[{Kuppens \emph{et~al.}(2010)Kuppens, Oravecz and
  Tuerlinckx}]{Kuppens2010}
Kuppens, P.; Oravecz, Z.; Tuerlinckx, F. (2010).
\newblock {Feelings change: Accounting for individual differences in the
  temporal dynamics of affect}.
\newblock \emph{Journal of Personality and Social Psychology} \textbf{99(6)},
  1042--1060.

\bibitem[{Lachanski and Pav(2017)}]{Lachanski2017}
Lachanski, M.; Pav, S. (2017).
\newblock Shy of the Character Limit:" Twitter Mood Predicts the Stock Market"
  Revisited.
\newblock \emph{Econ Journal Watch} \textbf{14(3)}, 302.

\bibitem[{Lodewyckx \emph{et~al.}(2011)Lodewyckx, Tuerlinckx, Kuppens, Allen
  and Sheeber}]{Lodewyckx2011}
Lodewyckx, T.; Tuerlinckx, F.; Kuppens, P.; Allen, N.~B.; Sheeber, L. (2011).
\newblock A hierarchical state space approach to affective dynamics.
\newblock \emph{Journal of Mathematical Psychology} \textbf{55(1)}, 68--83.

\bibitem[{Lorenz(2007)}]{Lorenz2007}
Lorenz, J. (2007).
\newblock Continuous opinion dynamics under bounded confidence: A survey.
\newblock \emph{International Journal of Modern Physics C} \textbf{18(12)},
  1819--1838.

\bibitem[{M\"as and Flache(2013)}]{Mas2013}
M\"as, M.; Flache, A. (2013).
\newblock Differentiation without Distancing. Explaining Bi-Polarization of
  Opinions without Negative Influence.
\newblock \emph{PLoS ONE} \textbf{8(11)}, e74516.

\bibitem[{Mitrovi{\'c} and Tadi{\'c}(2012)}]{Mitrovic2012}
Mitrovi{\'c}, M.; Tadi{\'c}, B. (2012).
\newblock Dynamics of bloggers' communities: Bipartite networks from
  empirical data and agent-based modeling.
\newblock \emph{Physica A}
  \textbf{391(21)}, 5264--5278.

\bibitem[{P{\'a}ez \emph{et~al.}(2015)P{\'a}ez, Rim{\'e}, Basabe, Wlodarczyk
  and Zumeta}]{Paez2015}
P{\'a}ez, D.; Rim{\'e}, B.; Basabe, N.; Wlodarczyk, A.; Zumeta, L. (2015).
\newblock Psychosocial effects of perceived emotional synchrony in collective
  gatherings.
\newblock \emph{Journal of Personality and Social Psychology} \textbf{108(5)},
  711.

\bibitem[{Rank \emph{et~al.}(2013)Rank, Skowron and Garcia}]{Rank2013}
Rank, S.; Skowron, M.; Garcia, D. (2013).
\newblock Dyads to groups: modeling interactions with affective dialog systems.
\newblock \emph{Int. J. Comput. Linguistics Res} \textbf{4(1)}, 22--37.

\bibitem[{Reisenzein(1983)}]{Reisenzein1983}
Reisenzein, R. (1983).
\newblock The Schachter theory of emotion: Two decades later.
\newblock \emph{Psychological Bulletin} \textbf{94(2)}, 239.

\bibitem[{Russell(1980)}]{Russell1980}
Russell, J.~A. (1980).
\newblock {A circumplex model of affect}.
\newblock \emph{Journal of Personality and Social Psychology} \textbf{39(6)},
  1161--1178.

\bibitem[{Sander \emph{et~al.}(2005)Sander, Grandjean and Scherer}]{Sander2005}
Sander, D.; Grandjean, D.; Scherer, K.~R. (2005).
\newblock A systems approach to appraisal mechanisms in emotion.
\newblock \emph{Neural Networks} \textbf{18(4)}, 317--352.

\bibitem[{Scherer(1984)}]{Scherer1984}
Scherer, K.~R. (1984).
\newblock On the nature and function of emotion: A component process approach.
\newblock \emph{Approaches to emotion} \textbf{2293}, 317.

\bibitem[{Scherer(2005)}]{Scherer2005}
Scherer, K.~R. (2005).
\newblock What are emotions? And how can they be measured?
\newblock \emph{Social Science Information} \textbf{44(4)}, 695--729.

\bibitem[{von Scheve and Salmella(2014)}]{vonScheve2014}
von Scheve, C.; Salmella, M. (2014).
\newblock \emph{Collective Emotions}.
\newblock OUP Oxford.

\bibitem[{Schwarz \emph{et~al.}(1991)Schwarz, Bless and Bohner}]{Schwarz1991}
Schwarz, N.; Bless, H.; Bohner, G. (1991).
\newblock Mood and persuasion: Affective states influence the processing of
  persuasive communications.
\newblock \emph{Advances in Experimental Social Psychology} \textbf{24},
  161--199.

\bibitem[{Schweitzer(2003)}]{Schweitzer2007}
Schweitzer, F. (2003).
\newblock \emph{Brownian Agents and Active Particles: Collective Dynamics in
  the Natural and Social Sciences}.
\newblock Berlin: Springer.

\bibitem[{Schweitzer(2018)}]{Schweitzer2018}
Schweitzer, F. (2018).
\newblock An agent-based framework of active matter with applications in
  biological and social systems.
\newblock \emph{European Journal of Physics} \textbf{40(1)}.

\bibitem[{Schweitzer and Garcia(2010)}]{Schweitzer2010a}
Schweitzer, F.; Garcia, D. (2010).
\newblock {An agent-based model of collective emotions in online communities}.
\newblock \emph{European Physical Journal B} \textbf{77(4)}, 533--545.
\newblock ISSN 1434-6028.

\bibitem[{Shin and Lorenz(2010)}]{Shin2010}
Shin, J.~K.; Lorenz, J. (2010).
\newblock Tipping diffusivity in information accumulation systems: more links,
  less consensus.
\newblock \emph{Journal of Statistical Mechanics: Theory and Experiment}
  \textbf{2010(06)}, P06005.

\bibitem[{Skowron \emph{et~al.}(2017)Skowron, Rank, Garcia and
  Ho{\l}yst}]{Skowron2017}
Skowron, M.; Rank, S.; Garcia, D.; Ho{\l}yst, J.~A. (2017).
\newblock Zooming in: studying collective emotions with interactive affective
  systems.
\newblock In: \emph{Cyberemotions}, Springer. pp. 279--304.

\bibitem[{Smith and Conrey(2007)}]{Smith2007}
Smith, E.~R.; Conrey, F.~R. (2007).
\newblock Agent-based modeling: A new approach for theory building in social
  psychology.
\newblock \emph{Personality and Social Psychology Review} \textbf{11(1)},
  87--104.

\bibitem[{Sobkowicz(2012)}]{Sobkowicz2012}
Sobkowicz, P. (2012).
\newblock Discrete Model of Opinion Changes Using Knowledge and Emotions as
  Control Variables.
\newblock \emph{PLoS ONE} \textbf{7(9)}, e44489.

\bibitem[{Sobkowicz(2013)}]{Sobkowicz2013}
Sobkowicz, P. (2013).
\newblock Quantitative Agent Based Model of User Behavior in an Internet
  Discussion Forum.
\newblock \emph{PLoS ONE} \textbf{8(12)}, e80524.

\bibitem[{Sobkowicz(2015)}]{Sobkowicz2015}
Sobkowicz, P. (2015).
\newblock Extremism without extremists: Deffuant model with emotions.
\newblock \emph{Frontiers in Physics} \textbf{3(17)}.

\bibitem[{Tadi{\'c} \emph{et~al.}(2013)Tadi{\'c}, Gligorijevi{\'c},
  Mitrovi{\'c} and {\v{S}}uvakov}]{Tadic2013}
Tadi{\'c}, B.; Gligorijevi{\'c}, V.; Mitrovi{\'c}, M.; {\v{S}}uvakov, M.
  (2013).
\newblock Co-evolutionary mechanisms of emotional bursts in online social
  dynamics and networks.
\newblock \emph{Entropy} \textbf{15(12)}, 5084--5120.

\bibitem[{Tadi{\'c} and {\v{S}}uvakov(2013)}]{Tadic2013b}
Tadi{\'c}, B.; {\v{S}}uvakov, M. (2013).
\newblock Can human-like bots control Collective mood: Agent-based simulations
  of online chats.
\newblock \emph{Journal of Statistical Mechanics: Theory and Experiment}
  \textbf{2013(10)}, P10014.

\bibitem[{Tadi{\'c} \emph{et~al.}(2017)Tadi{\'c}, {\v{S}}uvakov, Garcia and
  Schweitzer}]{Tadic2017}
Tadi{\'c}, B.; {\v{S}}uvakov, M.; Garcia, D.; Schweitzer, F. (2017).
\newblock Agent-based simulations of emotional dialogs in the online social
  network myspace.
\newblock In: \emph{Cyberemotions}, Springer. pp. 207--229.

\bibitem[{Von~Scheve and Ismer(2013)}]{vonScheve2013}
Von~Scheve, C.; Ismer, S. (2013).
\newblock Towards a theory of collective emotions.
\newblock \emph{Emotion Review} \textbf{5(4)}, 406--413.

\end{thebibliography}
\end{document}